\documentclass{aa}

\usepackage{graphicx}
\usepackage{txfonts}
\usepackage{color}
\usepackage{stfloats}
\usepackage{multirow}

\newcommand{\tablenotea}[1]{\parbox{7.7cm}{\indent \footnotesize{#1}}}

\newcommand{\bcsj}{Bull. Chem. Soc. Japan}

\newcommand{\chemrev}{Chem. Rev.}

\newcommand{\jcsft}{J. Chem. Soc. Faraday Trans.}

\newcommand{\jms}{J. Mol. Spectr.}

\newcommand{\jpca}{J. Phys. Chem. A}

\newcommand{\natastro}{Nat. Astron.}

\newcommand{\pnas}{PNAS}

\newcommand{\science}{Science}
\newcommand{\sadv}{Sci. Adv.}

\begin{document}

\title{Aromatic cycles are widespread in cold clouds\thanks{Based on observations carried out with the Yebes 40m telescope (projects 20A016, 21A006, and 22B023). The 40m radio telescope at Yebes Observatory is operated by the Spanish Geographic Institute (IGN; Ministerio de Transportes, Movilidad y Agenda Urbana).}}

\titlerunning{Aromatic cycles are widespread in cold clouds}
\authorrunning{Ag\'undez et al.}

\author{M.~Ag\'undez\inst{1}, N.~Marcelino\inst{2,3}, B.~Tercero\inst{2,3}, \and J.~Cernicharo\inst{1}}

\institute{
Instituto de F\'isica Fundamental, CSIC, Calle Serrano 123, E-28006 Madrid, Spain\\ \email{marcelino.agundez@csic.es} 
\and
Observatorio Astron\'omico Nacional, IGN, Calle Alfonso XII 3, E-28014 Madrid, Spain 
\and
Observatorio de Yebes, IGN, Cerro de la Palera s/n, E-19141 Yebes, Guadalajara, Spain
}

\date{Received; accepted}

 
\abstract
{We report the detection of large hydrocarbon cycles toward several cold dense clouds. We observed four sources (L1495B, Lupus-1A, L483, and L1527) in the Q band (31-50 GHz) using the Yebes\,40m radiotelescope. Using the line stack technique, we find statistically significant evidence of benzonitrile (C$_6$H$_5$CN) in L1495B, Lupus-1A, and L483 at levels of 31.8\,$\sigma$, 15.0\,$\sigma$, and 17.2\,$\sigma$, respectively, while there is no hint of C$_6$H$_5$CN in the fourth source, L1527. The column densities derived are in the range (1.7-3.8)\,$\times$\,10$^{11}$ cm$^{-2}$, which is somewhat below the value derived toward the cold dense cloud \mbox{TMC-1}. When we simultaneously analyze all the benzonitrile abundances derived toward cold clouds in this study and in the literature, a clear trend emerges in that the higher the abundance of HC$_7$N, the more abundant C$_6$H$_5$CN is. This indicates that aromatic cycles are especially favored in those interstellar clouds where long carbon chains are abundant, which suggests that the chemical processes that are responsible for the formation of linear carbon chains are also behind the synthesis of aromatic rings. We also searched for cycles other than benzonitrile, and found evidence of indene (C$_9$H$_8$), cyclopentadiene (C$_5$H$_6$), and 1-cyano cyclopentadiene (1-C$_5$H$_5$CN) at levels of 9.3\,$\sigma$, 7.5\,$\sigma$, and 8.4\,$\sigma$, respectively, toward L1495B, which shows the strongest signal from C$_6$H$_5$CN. The relative abundances between the various cycles detected in L1495B are consistent --within a factor of three-- with those previously found in \mbox{TMC-1}. It is therefore likely that not only C$_6$H$_5$CN but also other large aromatic cycles are abundant in clouds rich in carbon chains.}


\maketitle

\section{Introduction}

Cold dense clouds with temperatures of around 10 K and volume densities of a few 10$^4$ molecules of H$_2$ per cm$^3$ \citep{Bergin2007} are known to contain a rich variety of molecules, most of which are organic in nature. The chemical composition of this kind of environment has long been thought to be dominated by linear unsaturated carbon chains with a backbone of alternating triple and single bonds, such as polyynes and cyanopolyynes. Chemical models have been developed to explain the formation of molecules in these clouds, and they are reasonably successful in accounting for the abundances of the carbon chains observed in \mbox{TMC-1}, which is the paradigm of a cold dense cloud rich in carbon chains \citep{Agundez2013}. The quest for chemical complexity in these clouds has for years been focused on the search for increasingly long linear chains, the largest one found to date being HC$_{11}$N \citep{Loomis2021}.

However, in recent years we have witnessed a paradigm shift concerning the chemistry of cold dense clouds. Large hydrocarbon cycles, some of them aromatic in character, have been discovered in \mbox{TMC-1} by means of radioastronomical techniques. The first such molecule detected was benzonitrile (C$_6$H$_5$CN; \citealt{McGuire2018}), which is the CN derivative of benzene (C$_6$H$_6$) and is thought to be a proxy for the radioastronomically invisible aromatic molecule C$_6$H$_6$ \citep{Cooke2020}. The CCH derivative of benzene (C$_6$H$_5$CCH) has also been identified toward \mbox{TMC-1} \citep{Cernicharo2021c,Loru2023}, and the six-membered ring benzyne (C$_6$H$_4$) has also been observed \citep{Cernicharo2021b}.

The five-membered ring cyclopentadiene (C$_5$H$_6$), which is nonaromatic but moderately polar, has been directly observed toward \mbox{TMC-1} \citep{Cernicharo2021a}. In addition, the two CN isomeric derivatives \citep{McCarthy2021,Lee2021} and the two CCH substituted isomers \citep{Cernicharo2021c} have been detected. Moreover, fulvenallene (C$_5$H$_4$CCH$_2$), which is also a derivative of cyclopentadiene, has been observed as well \citep{Cernicharo2022}.

In addition to single cycles, hydrocarbons composed of fused rings have also been detected toward \mbox{TMC-1}. The two simplest members of the family of polycyclic aromatic hydrocarbons, indene (C$_9$H$_8$) and naphthalene (C$_{10}$H$_8$), have been detected, either directly or indirectly. The nonzero, albeit small dipole moment of indene permitted its detection \citep{Cernicharo2021a,Burkhardt2021a}, while one of the CN-substituted isomers of indene was also detected \citep{Sita2022}. Naphthalene is nonpolar and therefore cannot be observed through radioastronomical techniques. Nevertheless, the two CN-substituted isomers have been observed \citep{McGuire2021} and, similarly to the case of benzene, can be considered as proxies for naphthalene itself.

The observation of large cycles in cold dense clouds is challenging and has for this reason been restricted to \mbox{TMC-1}. In a recent study, \cite{Burkhardt2021b} reported the detection of C$_6$H$_5$CN toward four additional clouds. There are still many open questions regarding the presence of large hydrocarbon cycles in cold dense clouds. For example, the circumstances under which their presence is favored and how they  are synthesized remain unknown. Here we report the detection of large cycles toward new sources, which allows us to build a larger statistical basis from which to understand the occurrence of large hydrocarbon cycles in cold interstellar clouds.

\section{Astronomical observations}

\begin{figure*}
\centering
\includegraphics[angle=0,width=\textwidth]{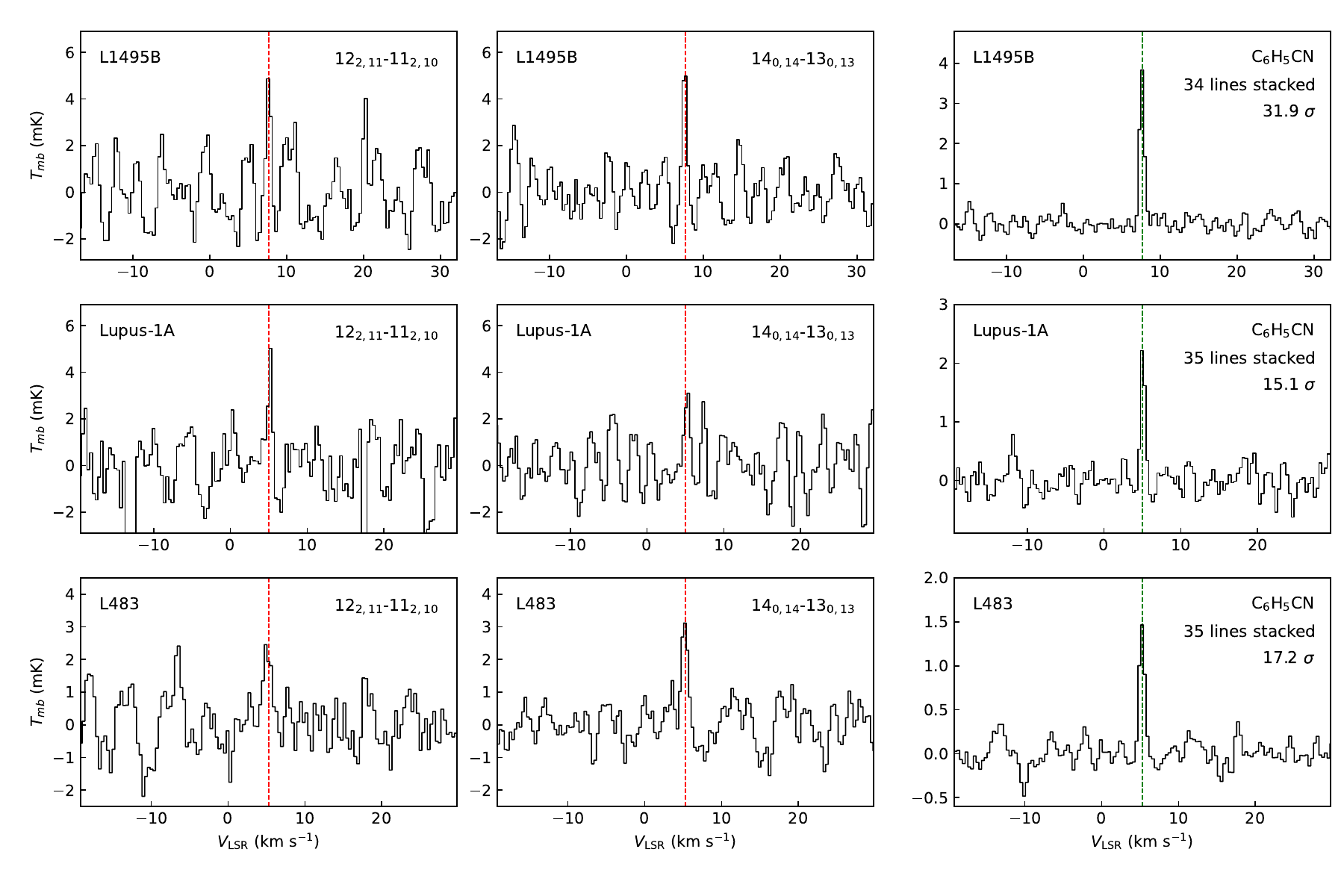}
\caption{Two individual lines of C$_6$H$_5$CN (left and middle panels) and the stacked spectrum of C$_6$H$_5$CN (right panels) for L1495B (top row), Lupus-1A (middle row), and L483 (bottom row), using the 13$_{0,13}$-12$_{0,12}$ line at 32833.813 MHz as a reference. The number of stacked lines and the significance level of the line stack signal (in units of $\sigma$) are indicated in the right panels. The systemic velocity of each source is indicated by a vertical dotted line.} \label{fig:lines_c6h5cn}
\end{figure*}

We carried out observations of four cold dense clouds (L1495B, Lupus-1A, L483, and L1527) with the Yebes\,40m telescope in the frame of a line survey in the Q band and a search for negative ions (details and source coordinates are given in \citealt{Agundez2023}). The observations, which used the 7\,mm receiver described in \cite{Tercero2021}, were taken between July 2020 and February 2023. The data consist of a full spectrum across the Q band (31.08-49.52 GHz) in horizontal and vertical polarizations with a spectral resolution of 38 kHz.

The observations were carried out using the frequency-switching technique. The frequency throw adopted, which changed between different observing periods as a result of tests done at the 40m telescope, was 10.52 MHz for L1495B and Lupus-1A, 10 MHz and 10.52 MHz for L483, and 8 MHz for L1527. We used two spectral setups with the frequency of the local oscillator separated by 5 MHz to identify spurious signals. Total on-source telescope times in each polarization were 45 h for L1495B, 120 h for Lupus-1A, 103 h for L483, and 40 h for L1527. The intensity scale at the 40m telescope is antenna temperature, $T_A^*$, for which we estimate a calibration error of 10\,\%. The $T_A^*$ noise levels after averaging horizontal and vertical polarizations are in the range 0.8-2.6 mK for L1495B, 0.7-2.8 mK for Lupus-1A, 0.4-1.0 mK for L483, and 0.7-2.7 mK for L1527. We convert $T_A^*$ to $T_{mb}$ (main beam brightness temperature) by dividing by $B_{\rm eff}$/$F_{\rm eff}$, where for the 40m telescope in the Q band $B_{\rm eff}$\,=\,0.797\,$\exp$[$-$($\nu$(GHz)/71.1)$^2$] and $F_{\rm eff}$\,=\,0.97. The half power beam width (HPBW) can be fitted as a function of frequency as HPBW($''$)\,=\,1763/$\nu$(GHz).

\section{Detection by line stack} \label{sec:detection}

\begin{figure*}
\centering
\includegraphics[angle=0,width=\textwidth]{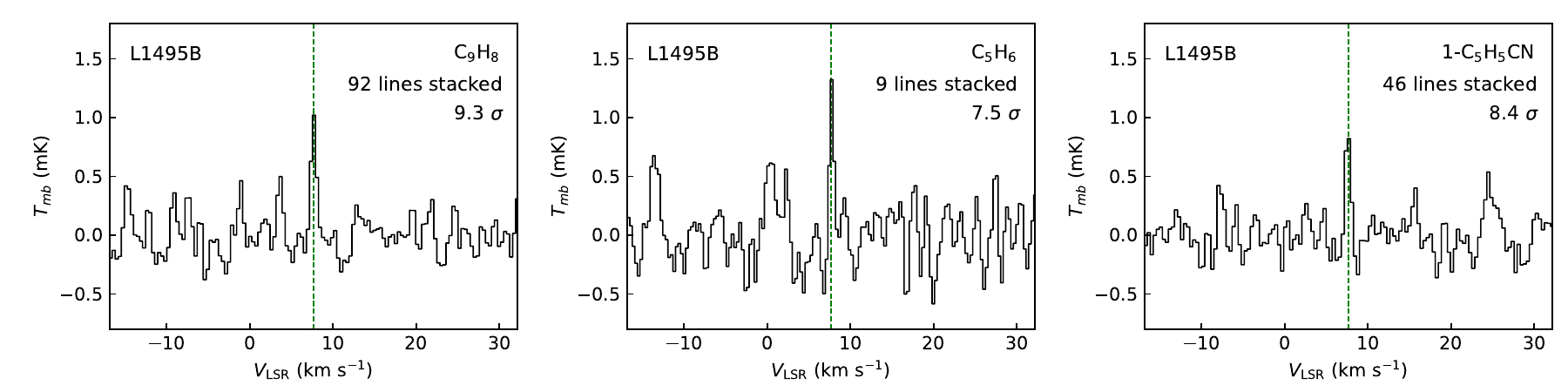}
\caption{Stacked spectra of L1495B for C$_9$H$_8$ (left panel), C$_5$H$_6$ (middle panel), and 1-C$_5$H$_5$CN (right panel), using the 14$_{0,14}$-13$_{0,13}$ line at 32528.329 MHz, the 3$_{1,2}$-2$_{2,1}$ line at 37453.992 MHz, and the 10$_{0,10}$-9$_{0,9}$ line at 33050.464 MHz as references, respectively. The number of stacked lines and the significance level of the line stack signal (in units of $\sigma$) is indicated in each panel. The systemic velocity of L1495B is indicated by a vertical dotted line.} \label{fig:lines_l1495b}
\end{figure*}

The Yebes\,40m data of the L1495B, Lupus-1A, L483, and L1527 were analyzed using the program CLASS of the GILDAS software \citep{Pety2005}\footnote{https://www.iram.fr/IRAMFR/GILDAS/}. While inspecting the data, we noticed the presence of several individual lines of C$_6$H$_5$CN in the spectrum of L1495B. The lines are weak, with $T_{mb}$\,$\sim$\,2-5 mK (see left panels in the top row in Fig.~\ref{fig:lines_c6h5cn}). We also searched for C$_6$H$_5$CN in the spectra of Lupus-1A and L483, although we only find a few lines, which are barely visible above the noise level (see left panels in the middle and bottom rows in Fig.~\ref{fig:lines_c6h5cn}), while many others are hidden within the noise. In the case of L1527, we see no obvious line of C$_6$H$_5$CN. Given that benzonitrile has many lines with similar intensities within the Q band, as seen in the very sensitive spectrum of \mbox{TMC-1} \citep{Cernicharo2021c}, we decided to use the line stack technique to evaluate whether or not there is spectral evidence of this molecule in the four aforementioned sources.

In order to carry out the line stack, we restricted to the most intense lines of C$_6$H$_5$CN in the Q band, which are those with quantum number $K_a$\,=\,0, 1, 2 in Figs.~C.1-C.4 of \cite{Cernicharo2021c} (35 lines in total). The hyperfine splitting of these lines is small (10-30 kHz). Lines with $K_a$\,$>$\,2 are weaker and have significant hyperfine splitting, which further dilutes the intensity among the different hyperfine components (see spectrum of \mbox{TMC-1} in \citealt{Cernicharo2021c}). The line frequencies and strengths of benzonitrile were computed from the rotational constants and dipole moments reported by \cite{Wohlfart2008}. We selected a spectral region of $\pm$\,30 km s$^{-1}$ around each line and cleaned the channels belonging to spectral features other than the C$_6$H$_5$CN target line; that is, lines from other molecules, negative frequency-switching artifacts, or spurious signals. In order to take into account the different expected intensities of the lines under consideration, we computed the spectrum under local thermodynamic equilibrium (LTE) at a temperature of 10 K. We note that the only determination of the rotational temperature of C$_6$H$_5$CN reported in the literature, which corresponds to \mbox{TMC-1} \citep{Cernicharo2021c}, yields a value of 9 K, which is equal to the gas kinetic temperature in this source \citep{Agundez2023}. Benzonitrile is therefore likely to be thermalized in cold dense clouds. The gas kinetic temperature is around 10 K in the four studied sources \citep{Agundez2023}, and therefore we adopt this round value here for simplicity. We arbitrarily selected the 13$_{0,13}$-12$_{0,12}$ line at 32833.813 MHz as a reference and used the calculated brightness temperatures of the LTE spectrum to compute relative intensities with respect to the reference line. We then scaled the spectrum of each line by the inverse of the theoretical relative intensity.

Once we had each spectrum cleaned from signals other than the target C$_6$H$_5$CN line and conveniently scaled according to the expected relative intensity, we fitted a baseline to a polynomial of degree 10 (the frequency-switching technique produces baseline ripples) and subtracted it. We then stacked all spectra aligned in the velocity scale using the inverse square of the rms noise level  as weight. The line-stacking procedure used here, which is similar to that employed to map C$_6$H$_5$CN in \mbox{TMC-1} \citep{Cernicharo2023}, has two benefits. On the one hand, it controls the absolute scale of the intensity, because all lines are stacked with the same intensity as that of the reference line. Therefore, the intensity of the signal after the line stack can be assigned to the reference line, which permits a correct evaluation of the column density. On the other hand, lines contribute to the signal according to their expected relative intensities. That is, the weaker the line, the higher the factor by which the spectrum is multiplied, the higher the noise level, and the lower the weight. This way, we avoid dilution of the line stack signal due to the weak lines.

The results from the line stack of C$_6$H$_5$CN are shown in the right panels of Fig.~\ref{fig:lines_c6h5cn} for L1495B, Lupus-1A, and L483. In the three cases, there is a clear signal well above the noise level and precisely centered at the systemic velocity of each source: 7.67 km s$^{-1}$ for L1495B \citep{Cordiner2013,Agundez2023}, 5.00 km s$^{-1}$ for Lupus-1A \citep{Sakai2010}, and 5.30 km s$^{-1}$ for L483 \citep{Agundez2019}. We note that in the case of L1495B, we removed one line from the analysis because it overlaps with a spurious signal. The signal-to-noise ratio (S/N) is evaluated as $\int T_A^* dv$ / [rms $\times$ $\sqrt{\Delta v \times \delta v}$], where rms is the measured noise level of the stacked spectrum, $\delta v$ is the spectral resolution in velocity of the stacked spectrum, and $\int T_A^* dv$ and $\Delta v$ are the velocity-integrated intensity and full width at half maximum (FWHM), respectively, given by a Gaussian fit to the line profile. Given the limitations associated to the line-stacking procedure, we consider that a detection is statistically significant when the S/N is above 5\,$\sigma$. We carried out the same line-stack analysis to search for evidence of C$_6$H$_5$CN in L1527, although no statistically significant signal was detected. The nondetection of C$_6$H$_5$CN in L1527 allows us to impose a stringent upper limit to the column density of C$_6$H$_5$CN. This is interesting because L1527 is known to be rich in carbon chains \citep{Sakai2008}.

Motivated by the detection of benzonitrile in the three clouds L1495B, Lupus-1A, and L483, we searched for evidence of other large cycles in these sources. Guided by previous observations toward \mbox{TMC-1} \citep{Cernicharo2021a,Cernicharo2021c,McGuire2021}, we searched for cyclopentadiene and its CN and CCH derivatives, for indene, and for the CN derivatives of naphthalene. For all these molecules, we computed the LTE spectrum at 10 K,  arbitrarily selecting one
intense line for reference (see caption of Fig.~\ref{fig:lines_l1495b}), and evaluated the relative line intensities from the calculated brightness temperatures. We only included those lines that are calculated to have at least half the intensity of the reference line. The line frequencies and strengths were computed from the rotational constants and dipole moments reported in the literature: C$_5$H$_6$ \citep{Bogey1988,Laurie1956}, 1- and 2-C$_5$H$_5$CN \citep{Cernicharo2021c,Sakaizumi1987}, 1- and 2-C$_5$H$_5$CCH \citep{Cernicharo2021c}, C$_9$H$_8$ \citep{Cernicharo2021a,Caminati1993}, and 1- and 2-C$_{10}$H$_7$CN \citep{McNaughton2018}. We then carried out the line-stack analysis in exactly the same manner as for C$_6$H$_5$CN. We found a statistically significant line stack signal for indene, cyclopentadiene, and 1-C$_5$H$_5$CN in the L1495B cloud. The stacked spectra are shown in Fig.~\ref{fig:lines_l1495b}, where we also give the number of lines included in the line stack analysis and the S/N reached. The significance levels are well above the 5\,$\sigma$ threshold. In any case, we plan to perform new observations of L1495B to provide a more robust identification of these three cycles and to search for other cycles that have not  yet been detected. The detection of these three particular cycles in L1495B is consistent with (1) the fact that this is the cloud with the highest signal for C$_6$H$_5$CN among the sources studied here, and (2) the relative intensities of the different cycles observed in \mbox{TMC-1}. In this regard, it makes sense that the only derivative of cyclopentadiene detected in L1495B is 1-C$_5$H$_5$CN because this is the one showing the higher line intensities in \mbox{TMC-1} \citep{Cernicharo2021c}. We also note that in the case of the CN derivatives of naphthalene, the search would be better carried out at frequencies below 30 GHz because of the small rotational constants.

\section{Discussion}

\begin{table}
\small
\caption{Observed column densities.}
\label{table:cdmol}
\centering
\begin{tabular}{llrr}
\hline \hline
Molecule & Source & \multicolumn{1}{c}{$T_{\rm rot}$\,$^a$} & \multicolumn{1}{c}{$N$} \\
               &              & \multicolumn{1}{c}{(K)}                 & \multicolumn{1}{c}{(cm$^{-2}$)} \\
\hline
C$_6$H$_5$CN & L1495B & & 3.8\,$\times$\,10$^{11}$ \\
C$_6$H$_5$CN & Lupus-1A & & 2.1\,$\times$\,10$^{11}$ \\
C$_6$H$_5$CN & L483 & & 1.7\,$\times$\,10$^{11}$ \\
C$_6$H$_5$CN & L1527 & & $<$6.5\,$\times$\,10$^{10}$\,$^b$ \\
C$_9$H$_8$ & L1495B & & 1.1\,$\times$\,10$^{13}$ \\
C$_5$H$_6$ & L1495B & & 7.8\,$\times$\,10$^{12}$ \\
1-C$_5$H$_5$CN & L1495B & & 6.6\,$\times$\,10$^{10}$ \\
& & & \\
HC$_7$N & L1495B & 9.3\,$\pm$\,0.4 & (2.7\,$\pm$\,0.4)\,$\times$\,10$^{12}$ \\
HC$_7$N & Lupus-1A & 9.7\,$\pm$\,0.4 & (6.3\,$\pm$\,0.9)\,$\times$\,10$^{12}$ \\
HC$_7$N & L483 & 11.1\,$\pm$\,0.4 & (2.2\,$\pm$\,0.3)\,$\times$\,10$^{12}$ \\
HC$_7$N & L1527 & 17.0\,$\pm$\,1.2 & (7.4\,$\pm$\,1.2)\,$\times$\,10$^{11}$ \\
\hline
\end{tabular}
\tablenotea{\\
$^a$ The rotational temperature for the cycles detected through line stack is assumed to be 10 K in all sources.\\
$^b$ Upper limit calculated at the 5\,$\sigma$ level.
}
\end{table}

To compute column densities for the cycles detected through line stack, we assumed that the velocity-integrated intensity of the stacked line corresponds to the reference line and adopted a round value of 10 K for the rotational temperature. The column densities derived are given in Table~\ref{table:cdmol}. In the case of L1527, we derived a 5\,$\sigma$ upper limit to the column density of C$_6$H$_5$CN. 

Benzonitrile is the aromatic cycle that has been most widely observed in cold dense clouds. In addition to the detection in \mbox{TMC-1} \citep{McGuire2018,Cernicharo2021c}, it has also been observed in four other cold clouds \citep{Burkhardt2021b}, three clumps in the Serpens South star-forming region (named 1a, 1b, and 2; \citealt{Friesen2013}), and in the L1521F core, located in the Taurus Molecular Cloud region \citep{Crapsi2004}. In their study, \cite{Burkhardt2021b} find that the abundance of benzonitrile relative to HC$_7$N varies considerably between the Taurus and Serpens sources. More specifically, the abundance ratio C$_6$H$_5$CN/HC$_7$N is about four times higher in the Taurus clouds \mbox{TMC-1} and L1521F compared to the three Serpens clouds. Here we report the detection of benzonitrile in three additional cold clouds and the nondetection in a fourth one. Therefore, we now have a sizable sample of nine sources where benzonitrile has been detected, or at least searched for, which can be used to examine how benzonitrile behaves among different cold clouds. The nine clouds are similar in that they are rich in carbon chains. It is therefore interesting to evaluate whether the abundance of benzonitrile is linked to that of carbon chains. To this purpose, we calculated column densities for HC$_7$N in the four clouds studied here using our Yebes\,40m data and the rotation diagram method. The calculated column densities, together with the rotational temperatures derived, are given in Table~\ref{table:cdmol}.

\begin{figure}
\centering
\includegraphics[angle=0,width=\columnwidth]{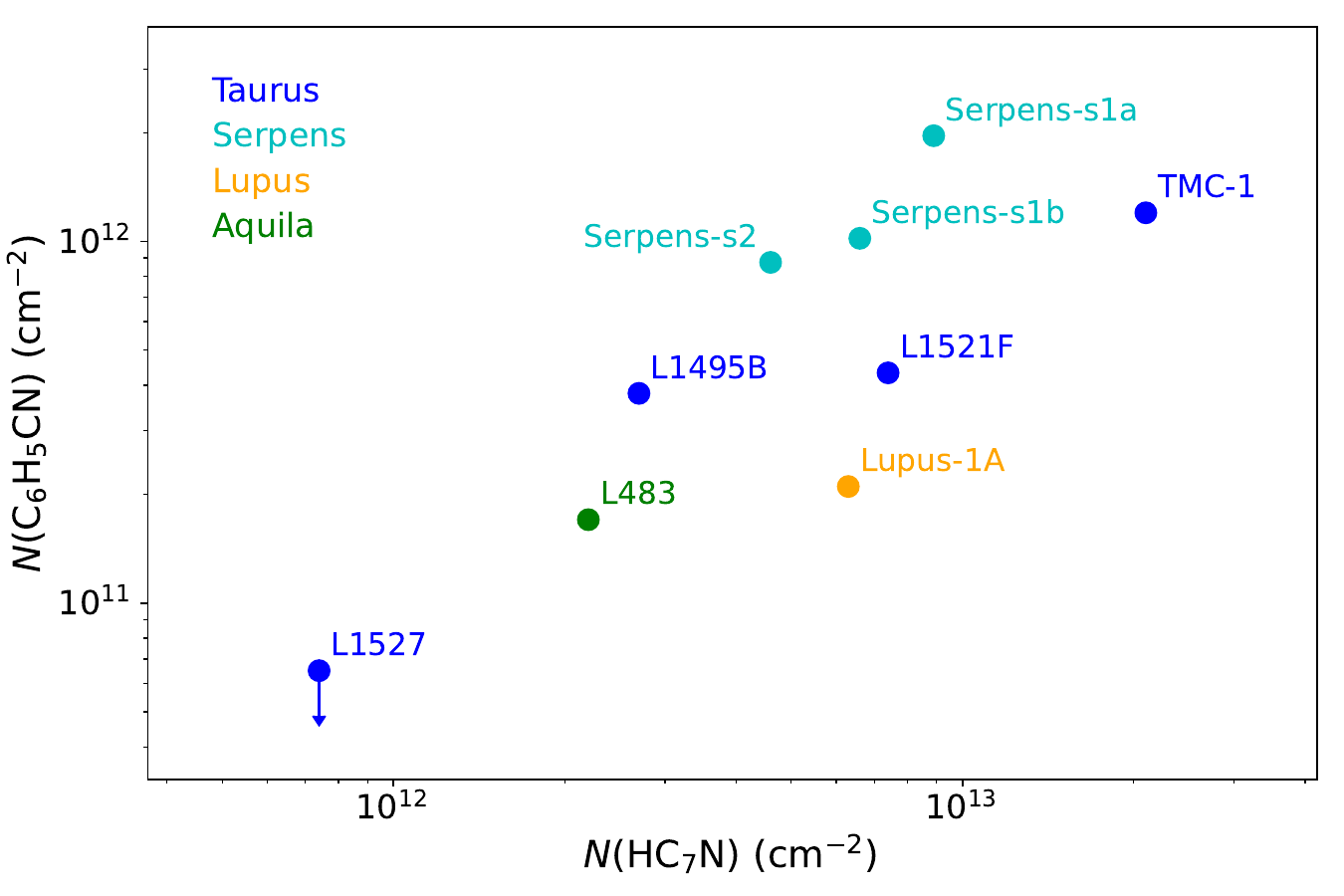}
\caption{Column density of C$_6$H$_5$CN versus column density of HC$_7$N for all the clouds where C$_6$H$_5$CN has been detected or searched for. The colors indicate different locations in the sky. Values were taken from \cite{Cabezas2022} and \cite{Cernicharo2021c} for \mbox{TMC-1}, from \cite{Friesen2013} and \cite{Burkhardt2021b} for the three clouds in Serpens and L1521F, and from the present work elsewhere.} \label{fig:c6h5cn_vs_hc7n}
\end{figure}

In Fig.~\ref{fig:c6h5cn_vs_hc7n} we represent the column densities of C$_6$H$_5$CN with respect to those of HC$_7$N. We can appreciate a very suggestive trend according to which the higher the column density of HC$_7$N, the more abundant C$_6$H$_5$CN is. The L1527 data point, although it only provides an upper limit to the column density of benzonitrile, brings valuable information supporting the positive correlation between aromatic cycles and linear chains. We can also see that the abundance of benzonitrile varies depending on the region where the cloud is located. For example, in Serpens-s1a, Serpens-s1b, L1521F, and Lupus-1A the column densities of HC$_7$N are similar, although the same is not true for C$_6$H$_5$CN, which is significantly more abundant in the two Serpens clouds than in Lupus-1A. Regarding the evolutionary status of the clouds, most sources in Fig.~\ref{fig:c6h5cn_vs_hc7n} do not contain a protostar, while L1521F hosts a very low-luminosity object \citep{Bourke2006}, and L483 and L1527 are known to contain a protostar \citep{Agundez2019,Sakai2008}. It is interesting to note that the two protostellar sources are the ones where benzonitrile is less abundant, although it is also true that these two clouds have lower  HC$_7$N column densities compared to the other sources. It therefore appears that aromatic cycles are favored in those clouds where carbon chains are abundant, and these clouds tend to be associated with early evolutionary stages along the process of star formation. The relationship between aromatic cycles and carbon chains is strengthened by the fact that the spatial distribution of C$_6$H$_5$CN in \mbox{TMC-1} coincides with that of carbon chains, such as HC$_3$N, HC$_5$N, HC$_7$N, and C$_6$H \citep{Cernicharo2023}.

Given that we detect several large cycles in L1495B, we are curious to know whether or not the relative abundances between the various cycles are similar to those found in \mbox{TMC-1}. Indene is the only species composed of fused rings that has been detected in L1495B. According to the values given in Table~\ref{table:cdmol}, indene has an abundance 29 times higher than that of benzonitrile. This value is on the order of that found in \mbox{TMC-1}, where the abundance ratio C$_9$H$_8$/C$_6$H$_5$CN is 13 \citep{Cernicharo2021a,Cernicharo2021c}. It is also interesting to note that the column density of indene in L1495B is of the same order as that of the single five-membered cycle cyclopentadiene (the C$_9$H$_8$/C$_5$H$_6$ ratio is 1.4; see Table~\ref{table:cdmol}), while the same is also found in \mbox{TMC-1}, where the C$_9$H$_8$/C$_5$H$_6$ abundance ratio is 1.3 \citep{Cernicharo2021a}. Finally, it is also interesting to examine the abundance ratio between a pure hydrocarbon cycle such as cyclopentadiene and its CN derivative. In L1495B, the abundance of the isomer 1-C$_5$H$_5$CN is 0.8\,\% of that of C$_5$H$_6$ (see Table~\ref{table:cdmol}), while in \mbox{TMC-1} this isomer is present with an abundance of 2.6\,\% of that of cyclopentadiene. In summary, the abundance ratios between different cycles in L1495B are consistent with those found in \mbox{TMC-1} within a factor of three.

%

The positive correlation between the abundance of C$_6$H$_5$CN and HC$_7$N found here, together with the coincident spatial distribution of benzonitrile and carbon chains found in \mbox{TMC-1} by \cite{Cernicharo2023}, suggest that there is a chemical link between aromatic cycles and carbon chains; that is, the same kind of chemistry that is responsible for the formation of linear chains is also behind the synthesis of aromatic rings. However, while the synthesis of carbon chains can be explained by current gas-phase chemical models of cold dense clouds (e.g., \citealt{Agundez2013}), it is not yet well understood how large hydrocarbon cycles are formed in these environments.

While it is nowadays generally accepted that aromatic rings are formed in situ in the cold clouds where they are detected, there is still debate over which processes drive the cyclization from small hydrocarbon species to single cycles such as cyclopentadiene and benzene. It has been proposed that benzene can be formed through ion-neutral reactions that lead to the C$_6$H$_7^+$ ion \citep{McEwan1999,Woods2011,Agundez2021}, although there are important uncertainties; to name one, the branching ratios of the dissociative recombination of the ion C$_6$H$_7^+$ that is thought to lead to benzene. Some neutral--neutral reactions have been studied and are promising routes to cyclopentadiene and benzene in cold clouds. For example, the reaction between 1,3-butadiene and the radical CH can yield cyclopentadiene \citep{He2020}, which upon further reaction with CH can produce benzene \citep{Caster2021}. Benzene can also be formed in the reaction between C$_2$H and 1,3-butadiene \citep{Jones2011}. However, when this chemical scheme (see Appendix D in \citealt{Cernicharo2021c}) is implemented in chemical models of cold dense clouds, the calculated abundances of benzonitrile and cyclopentadiene fall below the observed values \citep{Burkhardt2021b,Cernicharo2022}. There are still many reactions that need to be studied before understanding how cyclization occurs in cold clouds. A recent example is the study of the reaction between the propargyl and vinyl radicals by \cite{GarciadelaConcepcion2023}. Moreover, we also need to understand how single cycles such as cyclopentadiene and benzene convert into fused rings such as indene and naphthalene. In this context, \cite{Doddipatla2021} recently found that indene can be formed in the reaction of vinyl benzene with CH.

In addition to the challenge of explaining how aromatic cycles are formed in cold clouds, constraining how abundant and widespread they are is of use in evaluating their role in the physical and chemical evolution of the clouds. For example, cycles such as indene and naphthalene are little polar and have large sizes, which makes them strong candidates to be deposited as ices on dust grains. In a recent study, \cite{Mate2023} find that the abundance of indene observed in \mbox{TMC-1} implies that it should be present on ices with an exceedingly large abundance, accounting for up to 2\,\% of the carbon, which seems unrealistically high. It is clear that we still need to understand the interplay between gas and dust regarding these very big cycles. Astronomical observations able to constrain the type of region where aromatic rings are present and their abundances, combined with laboratory and theoretical studies, will eventually allow us to obtain a good understanding of the chemistry and physics of large aromatic rings in cold clouds.

\section{Conclusions}

We searched for large hydrocarbon cycles in four cold dense clouds. Using the line stack technique, we detected benzonitrile in three sources (L1495B, Lupus-1A, and L483), while it was not detected toward the fourth source (L1527). In addition, we detected three other cycles toward L1495B, namely indene, cyclopentadiene, and 1-cyano cyclopentadiene. By analyzing the abundances of benzonitrile derived here, together with values derived in the literature toward other sources, we find a positive correlation between the column density of C$_6$H$_5$CN and that of HC$_7$N. This result indicates that aromatic rings are favored in those regions where long carbon chains are abundant. The various cycles are likely to have similar relative abundances in cold dense clouds, as suggested by the comparison between L1495B and \mbox{TMC-1}.

\begin{acknowledgements}

We acknowledge funding support from Spanish Ministerio de Ciencia e Innovaci\'on through grants PID2019-106110GB-I00, PID2019-107115GB-C21, and PID2019-106235GB-I00. We thank the anonymous referee for a report that helped to improve this paper.

\end{acknowledgements}


\begin{thebibliography}{}

\bibitem[Ag\'undez \& Wakelam(2013)]{Agundez2013} Ag\'undez, M. \& Wakelam, V. 2013, \chemrev, 113, 8710
\bibitem[Ag\'undez et al.(2019)]{Agundez2019} Ag\'undez, M., Marcelino, N., Cernicharo, J., et al. 2019, \aap, 625, A147
\bibitem[Ag\'undez et al.(2021)]{Agundez2021} Ag\'undez, M., Cabezas, C., Tercero, B., et al. 2021, \aap, 647, L10 
\bibitem[Ag\'undez et al.(2023)]{Agundez2023} Ag\'undez, M., Marcelino, N., Tercero, B., et al. 2023, \aap, in press; DOI: 10.1051/0004-6361/202347077; arXiv:2307.04487
\bibitem[Bergin \& Tafalla(2007)]{Bergin2007} Bergin, E. A. \& Tafalla, M. 2007, \araa, 45, 339
\bibitem[Bogey et al.(1988)]{Bogey1988} Bogey, M., Demuynck, C. \& Destombes, J. L. 1988, \jms, 132, 277
\bibitem[Bourke et al.(2006)]{Bourke2006} Bourke, T. L., Myers, P. C., Evans II, N. J., et al. 2006, \apj, 649, L37
\bibitem[Burkhardt et al.(2021a)]{Burkhardt2021a} Burkhardt, A. M., Lee, K. L. K., Changala, P. B., et al. 2021a, \apj, 913, L18
\bibitem[Burkhardt et al.(2021b)]{Burkhardt2021b} Burkhardt, A. M., Loomis, R. A., Shingledecker, C. N., et al. 2021b, \natastro, 5, 181 
\bibitem[Cabezas et al.(2022)]{Cabezas2022} Cabezas, C., Ag\'undez, M., Marcelino, N., et al. 2022, \aap, 659, L8 
\bibitem[Caminati(1993)]{Caminati1993} Caminati, W. 1993, \jcsft, 89, 4153
\bibitem[Caster et al.(2021)]{Caster2021} Caster, K. L., Selby, T. M., Osborn, D. L., et al. 2021, \jpca, 125, 6927
\bibitem[Cernicharo et al.(2021a)]{Cernicharo2021a} Cernicharo, J., Ag\'undez, M., Cabezas, C., et al. 2021a, \aap, 649, L15 
\bibitem[Cernicharo et al.(2021b)]{Cernicharo2021b} Cernicharo, J., Ag\'undez, M., Kaiser, R. I., et al. 2021b, \aap, 652, L9 
\bibitem[Cernicharo et al.(2021c)]{Cernicharo2021c} Cernicharo, J., Ag\'undez, M., Kaiser, R. I., et al. 2021c, \aap, 655, L1 
\bibitem[Cernicharo et al.(2022)]{Cernicharo2022} Cernicharo, J., Fuentetaja, R., Ag\'undez, M., et al. 2022, \aap, 663, L9 
\bibitem[Cernicharo et al.(2023)]{Cernicharo2023} Cernicharo, J., Tercero, B., Marcelino, N., et al. 2023, \aap, 674, L4 
\bibitem[Cooke et al.(2020)]{Cooke2020} Cooke, I. R., Gupta, D., Messinger, J. P., \& Sims, I. R. 2020, \apj, 891, L41
\bibitem[Cordiner et al.(2013)]{Cordiner2013} Cordiner, M. A., Buckle, J. V., Wirstr\"om, E. S., et al. 2013, \apj, 770, 48
\bibitem[Crapsi et al.(2004)]{Crapsi2004} Crapsi, A., Caselli, P., Walmsley, C. M., et al. 2004, \aap, 420, 957
\bibitem[Doddipatla et al.(2021)]{Doddipatla2021} Doddipatla, S., Galimova, G. R., Wei, H., et al. 2021, \sadv, 7, eabd4044
\bibitem[Friesen et al.(2013)]{Friesen2013} Friesen, R. K., Medeiros, L., Schnee, S., et al. 2013, \mnras, 436, 1513
\bibitem[Garc\'ia de la Concepción et al.(2023)]{GarciadelaConcepcion2023} Garc\'ia de la Concepci\'on, J., Jim\'enez-Serra, I., Rivilla, V. M., et al. 2023, \aap, 673, A118
\bibitem[He et al. (2020)]{He2020} He, C., Zhao, L., Doddipatla, S., et al. 2020, Chem. Phys. Chem., 21, 1295
\bibitem[Jones et al.(2011)]{Jones2011} Jones, B. M., Zhang, F., Kaiser, R. I., et al. 2011, \pnas, 108, 452
\bibitem[Laurie(1956)]{Laurie1956} Laurie, V. M. 1956, \jcp, 24, 635
\bibitem[Lee et al.(2021)]{Lee2021} Lee, K. L. K., Changala, P. B., Loomis, R. A., et al. 2021, \apj, 910, L2
\bibitem[Loomis et al.(2021)]{Loomis2021} Loomis, R. A., Burkhardt, A. M., Shingledecker, C. N., et al. 2021, \natastro, 5, 188
\bibitem[Loru et al.(2023)]{Loru2023} Loru, D., Cabezas, C., Cernicharo, J., et al. 2023, \aap, in press; DOI: 10.1051/0004-6361/202347023
\bibitem[Mat\'e et al.(2023)]{Mate2023} Mat\'e, B., Tanarro, I., Tim\'on, V., et al. 2023, \mnras, 523, 5887
\bibitem[McCarthy et al.(2021)]{McCarthy2021} McCarthy, M. C., Lee, K. L. K., Loomis, R. A., et al. 2021, \natastro, 5, 176
\bibitem[McEwan et al.(1999)]{McEwan1999} McEwan, M. J., Scott, G. B. I., Adams, N. G., et al. 1999, \apj, 513, 287
\bibitem[McGuire et al.(2018)]{McGuire2018} McGuire, B. A., Burkhardt, A. M., Kalenskii, S., et al. 2018, \science, 359, 6372
\bibitem[McGuire et al.(2021)]{McGuire2021} McGuire, B. A., Loomis, R. A., Burkhardt, A. M. et al. 2021, \science, 371, 1265
\bibitem[McNaughton et al.(2018)]{McNaughton2018} McNaughton, D., Jahn, M. K., Travers, M. J., et al. 2018, \mnras, 476, 5268
\bibitem[Pety(2005)]{Pety2005} Pety, J. 2005, in SF2A-2005: Semaine de l'Astrophysique Francaise, ed. F. Casoli et al. (Les Ulis: EDP-Sciences), 721
\bibitem[Sakai et al.(2008)]{Sakai2008} Sakai, N., Sakai, T., Hirota, T., \& Yamamoto, S. 2008, \apj, 672, 371
\bibitem[Sakai et al.(2010)]{Sakai2010} Sakai, N., Shiino, T., Hirota, T., et al. 2010, \apj, 718, L49
\bibitem[Sakaizumi et al.(1987)]{Sakaizumi1987} Sakaizumi, T., Kikuchi, H., Ohashi, O., \&, Yamaguchi, I. 1987, \bcsj, 60, 3903
\bibitem[Sita et al.(2022)]{Sita2022} Sita, M. L., Changala, P. B., Xue, C., et al. 2022, \apj, 938, L12
\bibitem[Tercero et al.(2021)]{Tercero2021} Tercero, F., L\'opez-P\'erez, J. A., Gallego, J. D., et al. 2021, \aap, 645, A37
\bibitem[Wohlfart et al.(2008)]{Wohlfart2008} Wohlfart, K., Schnell, M., Grabow, J.-U., \& K\"upper, J. 2008, \jms, 247, 119
\bibitem[Woods(2011)]{Woods2011} Woods, P. M. 2011, in EAS Publication Series, 46, 235

\end{thebibliography}
\end{document}